# Monte Carlo study of a superlattice mixed spins under a crystal and external magnetic fields.

A. Mhirech and  L. Bahmad*

*Université Mohammed V, Faculté des Sciences, B.P. 1014, Rabat, Morocco.*

*Département de Physique, Laboratoire de Magnétisme et de la Physique des Hautes Energies.*

**Abstract:**

In this paper, we study the effect of[1] both an external and crystal magnetic fields for a superlattice with mixed spins (1/2, 1), using Monte Carlo simulations. The ground state phase diagrams, in the external magnetic field, coupling constant and the crystal magnetic field planes, were established. The effect of increasing temperatures has also been established showing the behavior of the magnetizations and susceptibilities, for fixed crystal field values. In particular, the obtained results as well as in the ground state phase diagrams, as for non null temperature are obtained for fixed external magnetic and crystal field values. On the other hand, the effect of the temperature on the hysteresis loops is presented showing that the decreasing temperature effect is to increase the absolute values of the corresponding coercitif magnetic field.

---

* corresponding author e-mail address: bahmad@fsr.ac.ma

**Introduction:**

Magnetic layered structures and superlattices have attracted significant attention recently by both theoretical and experimental researches [1] because of a wide array of fascinating properties. Besides, magnetic properties of Ising model with multispin interactions are of the interest theoretically and experimentally. Experimentally, the multispin interaction models were applied to several system physics such as classical fluids [2], binary alloys [3], lipid layers [4], rare gases [5] and metamagnets [6]. Furthermore, in recent years, there have been many theoretical study of two sublattice mixed spin $S_a$ and spin $S_b$ because they have less translational and their single counterparts, there are many phenomena in these systems that cannot be observed in the single-spin Ising model. Besides, the mixed systems are well adopted to study a certain type of ferrimagnetisms which are of great interest for the technological applications. Moreover, interesting and unusual phase diagrams may result in a superlattice. Many authors show that the superlattices exhibit the compensation behavior when their thicknesses are not very great [7,8]. Consequently, the study of the relation between the composition and the structure of these systems on the one hand, and their magnetic properties on the other hand, has become an active field of research [9,10].

Using the mean field theory and Monte Carlo simulations, the magnetic properties and layering transitions of several systems have been subject of intensive
recent studies [11-15]. In fact, the corresponding phase diagrams have been established as well as for pure and mixed systems.
On the other hand, several authors studied the effect of both external and crystal magnetic fields on magnetic behaviors and critical transition for magnetic films and superlattices with mixed spins [1619].

The aim of this work is to numerically study, using Monte Carlo simulations, the effect of both an external and a crystal magnetic fields on a superlattice with mixed spin ($S_a$=1/2,$S_b$=1). At first, we established the ground state phase diagrams, corresponding to the temperature $T=0$, in the planes: external magnetic field, coupling interaction between spins and the crystal magnetic field. After, we establish the superlattice magnetization and susceptibility behaviors temperature variation. The effect of external and crystal fields on superlattice magnetization is also calculated. Indeed, the obtained results in the ground state phase diagrams are also retrieved in the Monte Carlo simulations when decreasing temperature values. The behavior of the critical temperature versus the crystal magnetic field has also established for several system sizes.

**Model and method**:

In this work, we study a superlattice mixed Ising spins formed with alternating layers (see Fig. 1). The Hamiltonian governing this system is expressed as below:

$$\widetilde{H} = \widetilde{H}_1 + \widetilde{H}_2 + \widetilde{H}_3 \qquad (1)$$

with:

$$\widetilde{H}_1 = -J_h \sum_{<i,j>_n, <k,l>_n} (S_{bi}^n S_{aj}^n + S_{ak}^n S_{bl}^n) \qquad (2)$$

$$\widetilde{H}_2 = -J_p \sum_{<i>_n, <j>_n} [S_{ai}^n (S_{bi}^{n-1} + S_{bi}^{n+1}) + S_{bj}^n (S_{aj}^{n-1} + S_{aj}^{n+1})] \qquad (3)$$

$$\widetilde{H}_3 = \Delta \sum_{<i>_n} (S_{bi}^n)^2 + H \sum_{<i>_n} (S_{ai}^n + S_{bi}^n) , \qquad (4)$$

where $S_{ai} = \pm \frac{1}{2}$ and $S_{bi} = 0, \pm 1$ are the spin variables of the mixed system. $J_h$ is the coupling constant between the spins $S_{ai}$ and $S_{bi}$ in the same plane, $J_p$ stands for the coupling constant consecutive planes n and $n \pm 1$, see Fig.1. $\Delta$ and H are the crystal field and external magnetic field, respectively.

We performed Metropolis single-spin-flip Monte Carlo simulations on system sizes $L \times L \times D$, with L=16, 32, 64, 128 and D=4. We use periodic boundary conditions as well as in transverse D that in longitudinal $L \times L$ directions.

**Results and Discussion**

Figure 1 illustrates the drawing of the superlattice mixed spins (1/2,1) model. The mixed system is coupled by the constant $J_h$ between the spins $S_{ai}$ and $S_{bi}$ in the same plane. The spins of different layers are coupled by the constant $J_p$. The crystal field $\Delta$ is acting only on the spin-1, while the external magnetic field H is acting on all spins of the system. The superlattice we are studying is a system size formed with D square layers $L \times L$. We use periodic boundary conditions as well as in transverse D that in longitudinal $L \times L$ directions and L=16, 32, 64, 128 and D=4.

Preliminary results for several system sizes: D=4 and L=16, 32, 64, 128 have been performed for different studied parameters. The finite size scaling showed that the behavior of these parameters do not appreciably change when changing the system sizes. Consequently, in all the following, we will limit our calculations to the system size: D=4, L=4.

In order to outline the ground state phase diagram properties of the studied system, we show in Figs. 2a, 2b and 2c the different phases in the planes: (H, $\Delta$), ($J_p$, $\Delta$) and ($J_p$,H), respectively. Indeed, the only stable phases are (+1/2,+1), (-1/2,-1), (-1/2,0) and (+1/2,0), as it is shown in Fig.2a. While the only present phases in the plane ($J_p$, $\Delta$) are (-1/2,-1) and (-1/2,0), for the ferromagnetic case ($J_p > 0$), see Fig.2b. Besides, the same phases ((-1/2,-1) and (-1/2,0)) are found in the plane ($J_h$, $\Delta$), as it is shown in Fig.2c.

The behavior of the sub-lattice magnetizations $M_a$ and $M_b$ as a function of the temperature as illustrated in Figs.3a and 3b in the absence and the presence of the external magnetic field, respectively. Indeed, a critical temperature $T_c \approx 1.75$ is found in Fig.3a, whereas the critical temperature $T_c$ is never found in the presence of the external magnetic field, as it is illustrated in Fig.3b.

The temperature dependency of the sub-lattice susceptibilities per spin for $J_p$=1, $J_h$=1, L=32, $\Delta$=0 and H=0 are illustrated in Fig.4. These magnetic susceptibilities undergo a sharp pic at the critical temperature found in Fig.3a ($T_c \approx 1.75$).

The global superlattice magnetization versus the external magnetic field H is given in Figs.5a and 5b for $J_p$=1, $J_h$=1, $\Delta$=20 and $\Delta$=-20 in the case of two temperature values T=3 and T=0.01, respectively. Is is found that for a very low temperature value the total magnetization undergoes only one first order transition for the negative crystal magnetic field value $\Delta$=-20. Whereas, for the positive crystal magnetic field value $\Delta$=20, the total magnetization presents three first order transitions, see Fig.5a. On the other hand, for higher temperature value the total magnetization presents only second order transitions as well as for positive and negative crystal magnetic field values, see Fig.5b.

To highlight the second order transitions observed in Figure 5b, we plot in Fig. 6 the behavior of the superlattice mean susceptibility versus the external field H for $\Delta$=20 and $\Delta$=-20 in the case where $J_p$=1, $J_h$=1 and T=3. It is obvious that every second order transition in mean magnetization has resulted in a spike in global superlattice susceptibility. We observe three peaks in the case of $\Delta$=20 and a peak when $\Delta$=-20. Indeed, for $\Delta$=20 the susceptibility shows only one peak corresponding to the phases (+1/2,+1) and (-1/2,-1) shown in the ground state phase diagram. While for $\Delta$=-20, the three peaks presented by the magnetic susceptibility correspond to the shifting from the four phases: (+1/2,+1), (+1/2,0), (-1/2,0) and (-1/2,-1), respectively.

Finally, the effect of the temperature on the hysteresis loops for $J_p$=1, $J_h$=1, L=32 and $\Delta$=0 is presented in Fig.7. It is found that the decrease of the temperature favors the hysteresis loops, indeed it increases when decreasing the temperature.

**Conclusion**

In this paper we have used Monte Carlo simulations to study the effect of both an external and a crystal magnetic fields for a superlattice with mixed spins (1/2, 1).The ground state phase diagrams were established in the planes: external magnetic field, coupling interaction between spins and the crystal magnetic field, for T=0. In particular, the effect of increasing temperature has also been outlined in the behavior of the magnetizations and susceptibilities. The obtained results in the ground state phase diagrams are also retrieved in the Monte Carlo simulations when decreasing temperature values. Finally, the behavior of the critical temperature versus the crystal magnetic field has also been established for several system sizes and fixed parameter values.

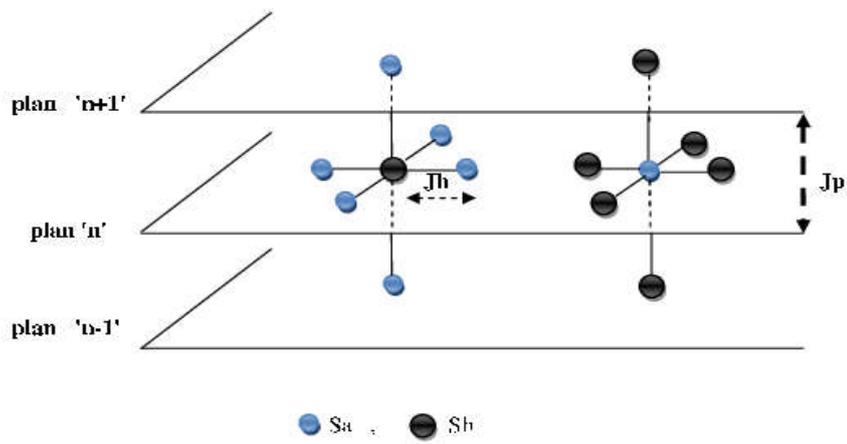

Figure 1: Schematic drawing of superlattice mixed spins (1/2,1) model.

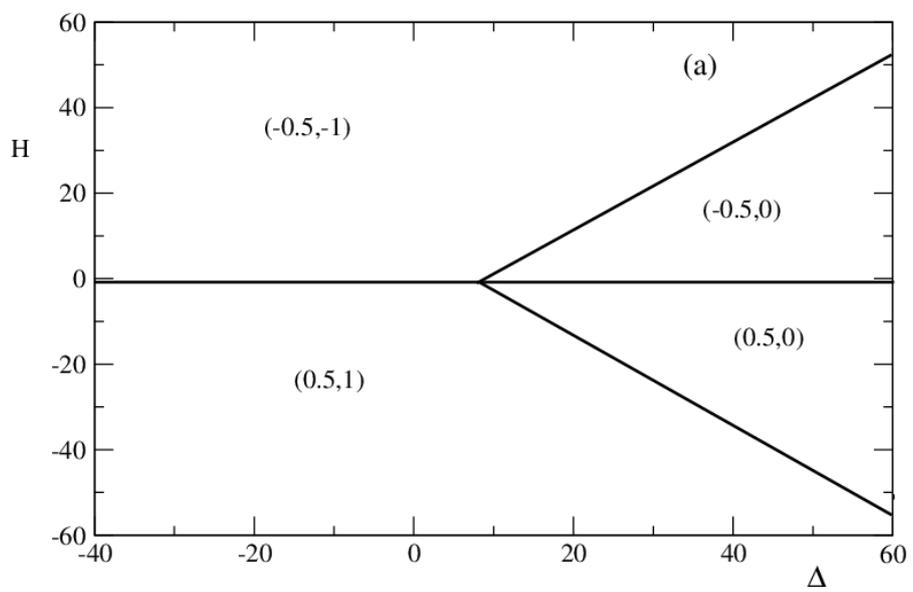

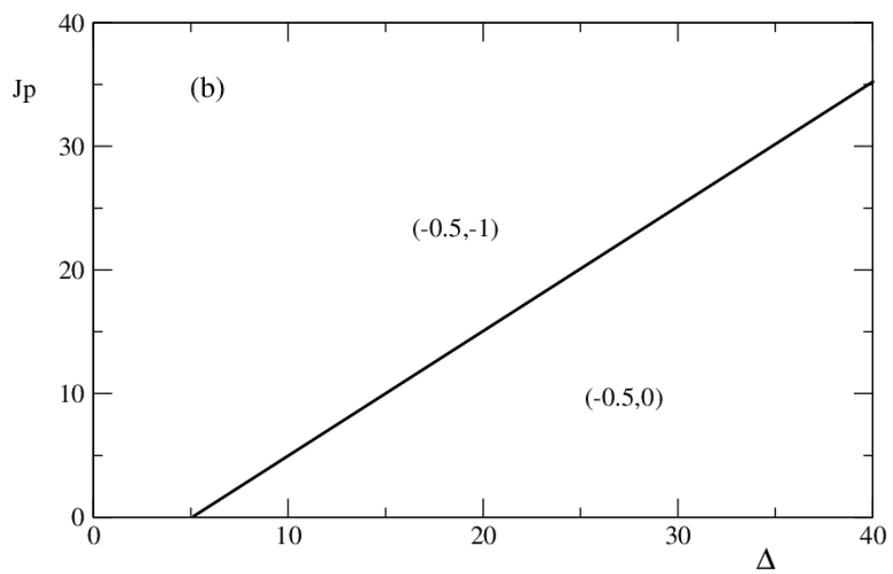

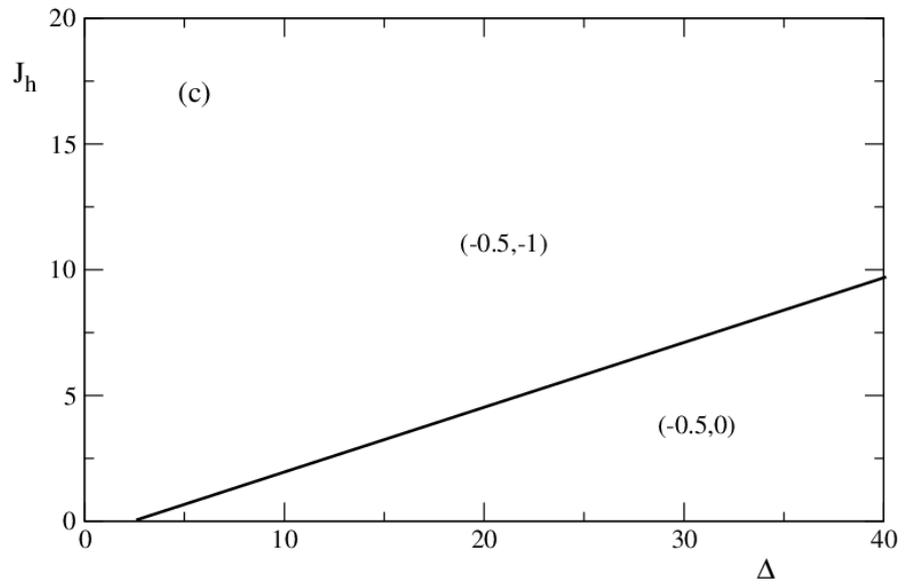

Figure 2: Ground state phase diagrams in the three follow cases: (a) (H, $\Delta$) plane with $J_p$=1 and $J_h$=1, (b) ($J_h$, $\Delta$) plane with H=1 and $J_p$=1 and (c) ($J_p$, $\Delta$) plane with H=1 and $J_h$=1. The system size is L=32.

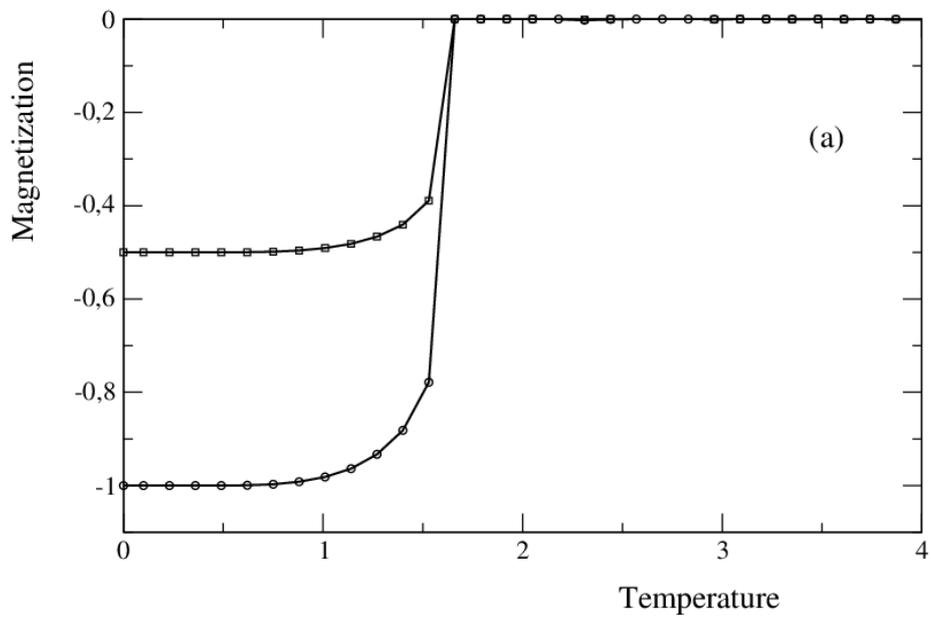

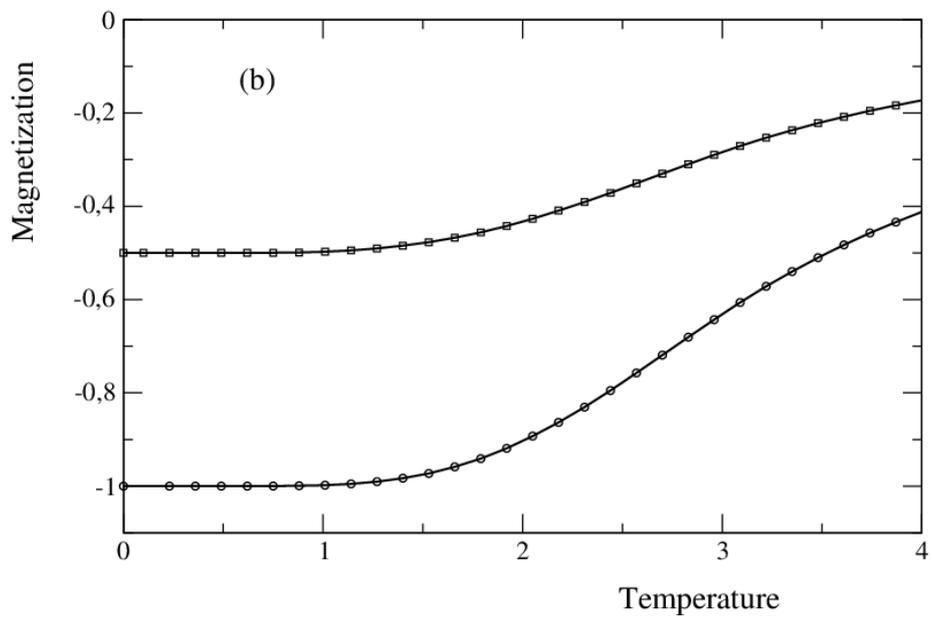

Figure 3: Temperature dependency of the sub-lattice magnetizations per spin for $J_p=1$, $J_h=1$, $\Delta=-10$ and $L=32$ in two following cases: (a) $H=1$, (b) $H=0$.

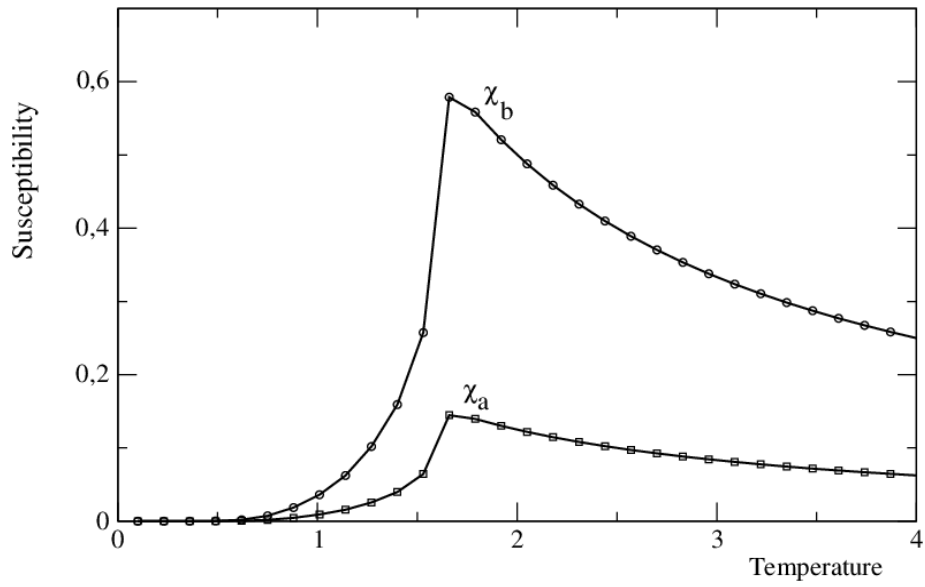

Figure 4: Temperature dependency of the sub-lattice susceptibilities per spin for $J_p=1$, $J_h=1$, $L=32$, $\Delta=0$ and $H=0$.

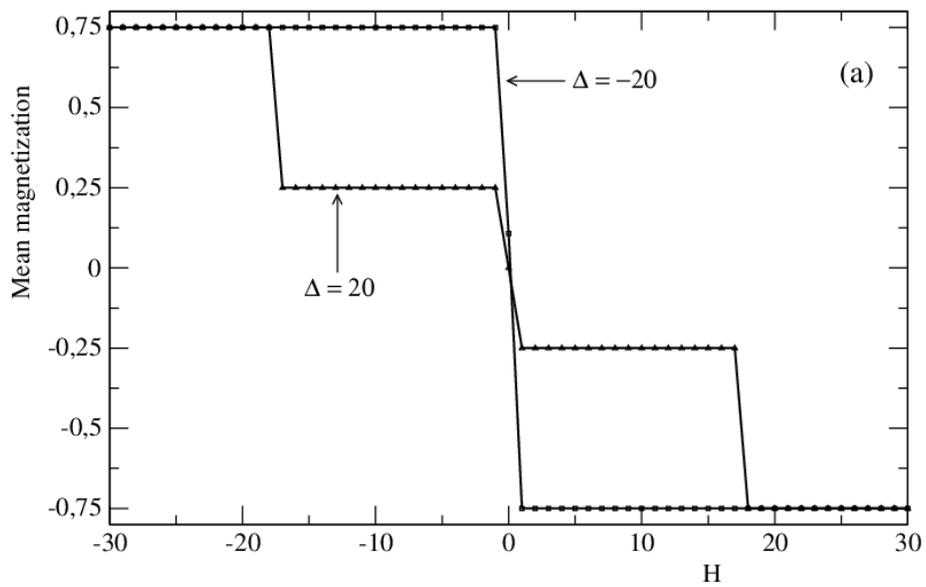

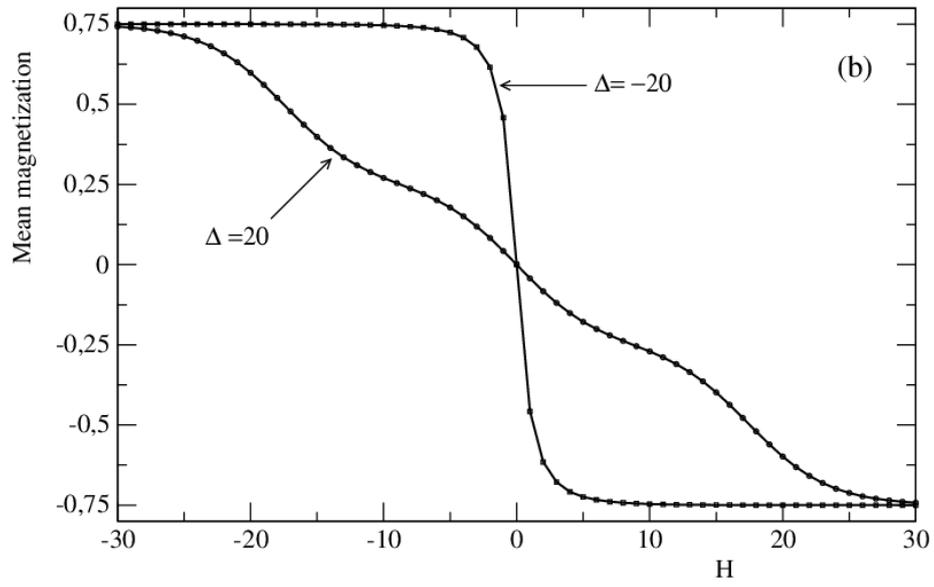

Figure 5: Total superlattice magnetization versus external field H for $J_p=1$, $J_h=1$, $\Delta=20$ and $\Delta=-20$ for two temperature values: (a) T=3 and (b) T=0.01.

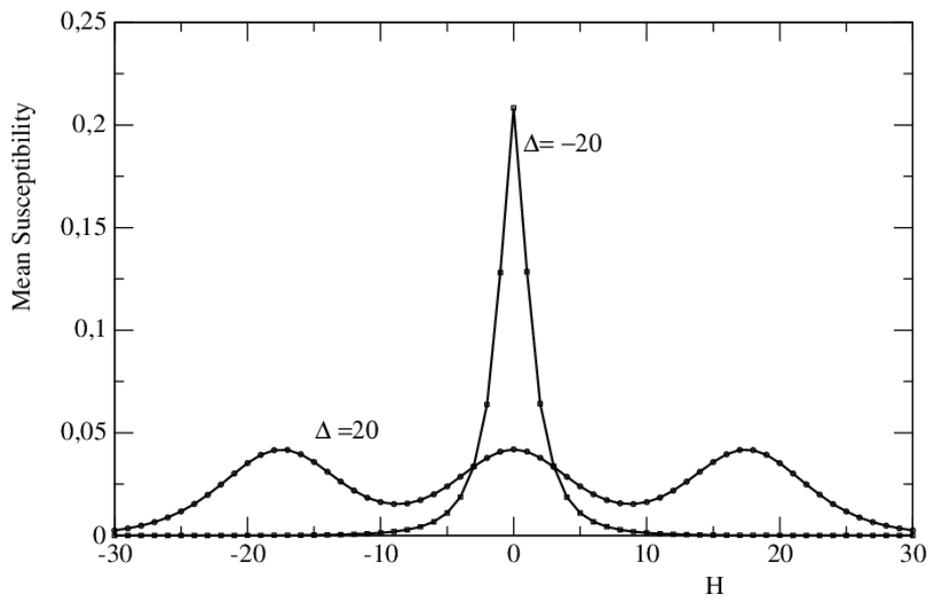

Figure 6: Variation of the superlattice mean susceptibility as a function of the external field H for $\Delta$ =20 and $\Delta$ =-20. $J_p$=1, $J_h$=1 and T=3.

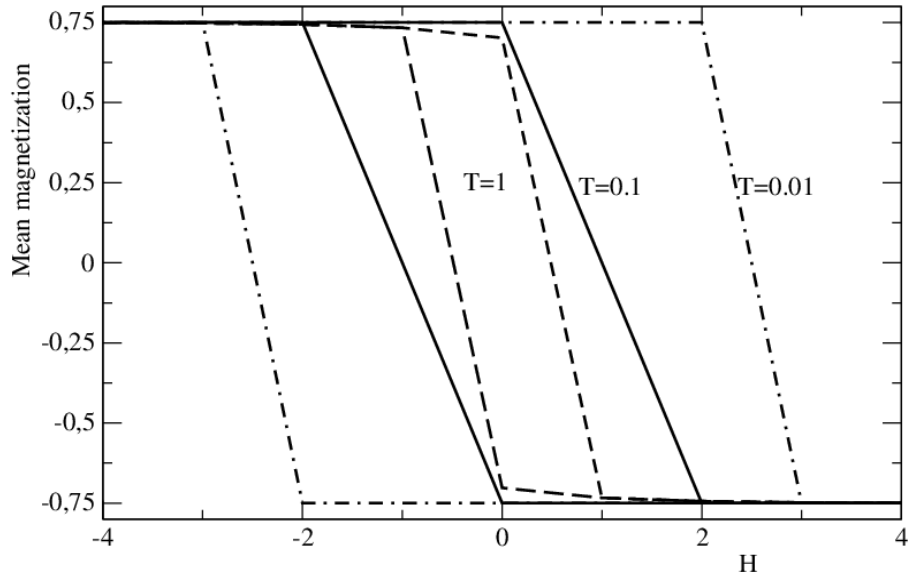

Figure 7: Temperature dependence of the hysteresis loops at $J_p$=1, $J_h$=1 , L=32 and $\Delta$=0.